\documentstyle[prl,aps,epsf]{revtex}
\newcommand{\bfm}[1]{\mbox{\boldmath${#1}$}}

\begin{document}
\draft
\twocolumn[\hsize\textwidth\columnwidth\hsize\csname
@twocolumnfalse\endcsname

\widetext
\title{ Addition energies of a Quantum Dot with harmonic electron-electron 
interactions }
\author {  Antimo Angelucci and  Arturo Tagliacozzo}
\address{ Dipartimento di Scienze Fisiche Universit\`a di Napoli, and INFM \\
Mostra d' Oltremare Pad. 19, I-80125 Napoli, Italy}
\maketitle
\begin{abstract}
We study a two dimensional electron system in a parabolic confining potential 
and constant magnetic field for the case of harmonic electron-electron 
interaction. We present analytic results for the electrochemical potential
versus magnetic field and discuss the effects of correlation in connection 
with the addition energy of a Quantum Dot with few electrons. 
\end{abstract}

]

\narrowtext

Recent achievements  of nanolithography  in semiconductor technology allow 
for the fabrication of devices in which a definite number of electrons are 
confined within two-dimensional islands of size as small as tenths of 
nanometers [Quantum Dots (QD)]\cite{dots}. In the last few years there has been 
growing interest in the study of these devices in view of improving our 
understanding of correlated electron systems. In fact, QD are unique with 
respect to other structures, e.g., macromolecules and clusters, because a dot 
can be connected to sources and/or a measuring apparatus via contacts. This 
possibility allows investigation of the system with probes changing the 
number of particles \cite{tunnel}. Indeed, quite recently 
Tarucha {\em et al.}\cite{tarucha} have measured the tunneling current in 
gated vertical quantum dots as a function of a magnetic field $B$ applied 
parallel to the current. In the Coulomb blockade regime and in presence of a 
very small voltage bias, the current shows a sequence of peaks that occur
whenever the gate voltage $V_g$ is proportional (via a voltage-to-energy 
conversion coefficient) to the chemical potential $\mu_N = E(N)-E(N-1)$ for 
adding one more particle to the dot. Here $E(N)$ is the ground-state (GS) 
energy of the dot once $N$ electrons are localized in it.  

General features of the current peaks are qualitatively reproduced by 
assuming that electrons are confined by a parabolic potential of 
frequency $ \omega_0$ and by adding the charging energy 
$E_{\rm ch}=V_0 N(N-1)/2$ to account for the electron-electron repulsion 
[Constant Interaction (CI) model]\cite{tarucha,mceuen}. The GS 
energy is obtained, for any value of $B$, by filling the lowest 
one-particle free harmonic oscillator levels with electrons of both spin.
In this model the observed increasing of the addition energy 
$\Delta_N = E(N+1) + E(N-1) - 2 E(N)$ for $N=N_p$, where
$N_p=2,6,10,20,...$, is easily related to the shell structure of the two 
dimensional harmonic potential 
spectrum, which leads to a marked increase of $\mu_N$ whenever a new shell 
is opened. The CI model then reproduces the oscillations of the current peaks 
with the field $B$ as well as their shift in pairs. Because the CI GS wave 
function has minimum total spin, its agreement with the experimental pattern 
also confirms that in some QD, like the one in Ref.\cite{tarucha}, the
effective $g$-factor $g_*$ can be very small\cite{wagner}. 
However, there are some evident features of the experiments that cannot be 
reproduced by the simple independent electron CI Hamiltonian.
Because the harmonic single-particle levels are equally spaced,
according to the CI model (at $B=0$) $\Delta_N$ should be 
constant ($\Delta_N=V_0$) within a unfilled shell. For the same reason  
one should then find 
$\Delta_{N_p} = V_0 + \hbar\omega_0$ for any $N_p$.  
While the experimental pattern in  Ref.\cite{tarucha} 
allows a clear identification of a shell structure for dots of less than 
$\approx 20$ electrons, it also clearly shows a smooth shrinkage of the spacing 
$\Delta_N$ within a unfilled shell with growing $N$, as it was
previously noticed 
in single electron capacitance spectroscopy\cite{ashoori2}. Then, one 
observes that the expected peaks at $\Delta_{N_p}$ decrease in height 
each time a new shell opens. In general, as 
discussed by Schmidt {\em et al}.\cite{schmidt}, fittings of capacitance data 
within the CI model indicate that the confinement strength of the potential 
appears to decrease rapidly with increasing energy. Question arises 
whether such an effect can be attributed to electron-electron correlation. 

In this letter we consider the model of harmonic interactions (HI) between $N$ 
electrons\cite{green} in a parabolic confining potential. Although there 
is tenuous  justification for this model interaction, when the
magnetic field is not too small and the electron number not too large
we show that it embodies correlation effects which correctly reproduce 
interesting experimental features. 
The model is well known and its popularity relies on the fact 
that for fully spin polarized electrons the Laughlin trial wave function, 
which is successful in describing the Coulomb gas in a quantizing 
magnetic field,
has a form similar to one of its eigenstates\cite{girvin}. Successively, 
the model has been discussed, again in the fully polarized spin sector, in 
connection with the QD problem in a large magnetic field orthogonal to the dot 
plane\cite{payne}. In view of the finding $g_*\approx 0$, results about the 
maximum spin sector are not relevant to the present discussion. We present 
some details of the solution because little information can be found in 
literature about the GS wave function in an arbitrary spin sector. 

We do not expect the HI model useful in discussing the region of small magnetic 
field. In fact, there are evidences\cite{tarucha} that in this case the spin 
sector of the QD GS obeys Hund's rule, whereas the GS of the HI model at $B=0$ 
has necessarily lowest spin\cite{lieb}. Instead, for higher magnetic field 
our results closely reproduce features of the experimental data of 
Ref.\cite{tarucha} and are summarized in Fig.~(1). The narrowing, compared 
with the CI model, of the distance between current peaks is transparent.
The shift, at fixed $V_0$, of the orbital angular momentum transitions which 
take place with increasing $B$ to lower values of the magnetic field also
clearly appears in our results. Then, close to the point where the
oscillations in the peak location drop, for large enough $N$, the peaks making 
up a pair have an  intriguing ``out of phase'' behaviour, similar to the
experimental pattern.  

The HI Hamiltonian in a constant magnetic field $B$ is 
$H_B = H_0 + \hbar {\omega}_c(L_3 +g_*S_3)/2$, where 
$\hbar {L}_3=\sum_{i=1}^N (x_{1,i}p_{2,i}-x_{2,i}p_{1,i})$ and 
$S_3=\sum_{i=1}^N S_{3,i}$ are the third component of the total angular 
momentum and total spin operators, respectively, 
${\bfm r}_i=(x_{1,i},x_{2,i})$ and 
${\bfm p}_i=-i\hbar\partial/\partial {\bfm r}_i$  are 
the coordinates and momentum of the $i$-th electron, ${\omega}_c = e B/(m_*c)$ 
is the cyclotron frequency, and 
\begin{eqnarray} 
H_0 = & - & {\hbar^2\over{2m_*}} \sum_{i=1}^N 
{{\partial}\over{\partial {\bfm r}_i}} \cdot 
{{\partial}\over{\partial {\bfm r}_i}} 
+ m_*{{\omega^2}\over2} \sum_{i=1}^N \vert {\bfm r}_i\vert ^2 
\nonumber\\
& + & \sum_{1\le i<j\le N}(V_0-
{U\over2} \vert {\bfm r}_i-{\bfm r}_j\vert^2). 
\eqnum{1} 
\label{H2} 
\end{eqnarray} 
Here $m_*$, $g_*$ denote effective parameters and for the time being we set 
$m_*=e=\hbar=1$. The frequency $\omega^2=\omega_0^2 +{\omega}_c^2/4$ enters 
$H_0$ upon choosing the gauge ${\bfm A}=(B/2)(-x_{2},x_{1})$ for the vector 
potential and represents the frequency of the effective parabolic confining 
potential, whereas $U\ge0$ is the strength of the interaction. For $U=0$ then 
$H_B$ reduces to the CI model. Introducing 
$\Lambda_{ij} = \Omega^2 \delta_{ij}+U{\cal J}_{ij}$, where  
$\Omega^2=\omega^2-NU$ and ${\cal J}_{ij}$ denotes the matrix with all 
unit entries, (i.e., ${\cal J}_{ij}=1$ $\forall\,i,j$), the potential energy 
entering Eq.~(\ref{H2}) can be compactly written  as 
$V = \sum_{i,j=1}^N\Lambda_{ij} {\bfm r}_i\cdot{\bfm r}_j/2$, so that $H_0$ 
is bound from below if $\Lambda_{ij}$ is positive definite. 
The symmetric matrix $\Lambda_{ij}$ is diagonalized by any unitary matrix 
${\cal U}_{i \nu}$ satisfying to 
$\sum_{i=1}^N{\cal U}_{i \nu} = 0$, $\forall\nu\ne N$, and 
${\cal U}_{i N} = N^{-{1\over2}}$, $\forall i$. Its eigenvalues 
$\lambda_1=\ldots=\lambda_{N-1}=\Omega^2$,  $\lambda_N = \omega^2$ 
are readily evaluated, so that positivity is ensured whenever 
$\omega^2>NU$. Hence, if we limit the discussion to $N$ not too large, 
the unphysical feature of dealing with an interaction unbounded at 
large distances is compensated by the presence of the confining 
potential. We then introduce normal coordinates
${\bfm y}_{\nu} = \sum_{i=1}^N {\cal U}^{\dagger}_{\nu i} {\bfm r}_i$. The 
Laplacian 
$\Delta = \sum_{i=1}^N {{\partial}/{\partial {\bfm r}_i}}\cdot 
{{\partial}/{\partial {\bfm r}_i}}$, the angular momentum ${L}_3$, and 
the operator $R^2 = \sum_{i=1}^N \vert {\bfm r}_i\vert^2$ are invariant under 
unitary transformations and from the invariance of $R^2$ one gets the simple 
but key identity
$\sum_{{\nu}=1}^{N-1} \vert {\bfm y}_{\nu}\vert^2 = 
\sum_{ i<j } \vert {\bfm r}_i-{\bfm r}_j\vert^2/N$. 
In the new basis the equation $H_B\Psi=E\Psi$ is immediately solved, because 
the key identity allows to write $H_0$ as a sum of separated harmonic 
oscillators of frequencies $\sqrt{\lambda_{\nu}}$. However, the 
straightforward normal-mode approach is of no much practical help, because the 
main problem one faces is to account for the identity of the particles. 
 Eigenfunctions of $H_B$ with definite symmetry 
under particle permutations can be factorized in the form 
$\Psi=\Psi_{\rm cm}\Psi_r$, in the usual way\cite{firsov}, 
where $\Psi_{\rm cm}$ is the completely symmetric harmonic wave 
function of the center of mass (c.m.) coordinate 
${\bfm r} = N^{-{1\over 2}} {\bfm y}_N$, 
with energy $E_{\rm cm}$ and angular momentum $L_3^{\rm cm}$, 
while the relative motion wave function $\Psi_r$ 
must be solution of the equation 
\begin{eqnarray}
- {1\over2} \Delta \Psi_r & + & {\Omega^2\over{2N}} \sum_{i<j} 
\vert {\bfm r}_i-{\bfm r}_j\vert^2 \Psi_r 
+ {{\omega}_c\over2}(L_3 +g_*S_3) \Psi_r 
\nonumber\\
& = & (E-E_{\rm ch}-E_{\rm cm}-{{\omega}_c\over2}L_3^{\rm cm})\Psi_r, 
\eqnum{2} 
\label{calogero_mod}
\end{eqnarray} 
of same symmetry as $\Psi$ and of zero total linear momentum 
${\bfm P} = \sum_{i=1}^N {\bfm p}_i$. 
\begin{figure}[m]
\null\vskip -0.truecm
\epsfxsize=7.7cm\epsfbox{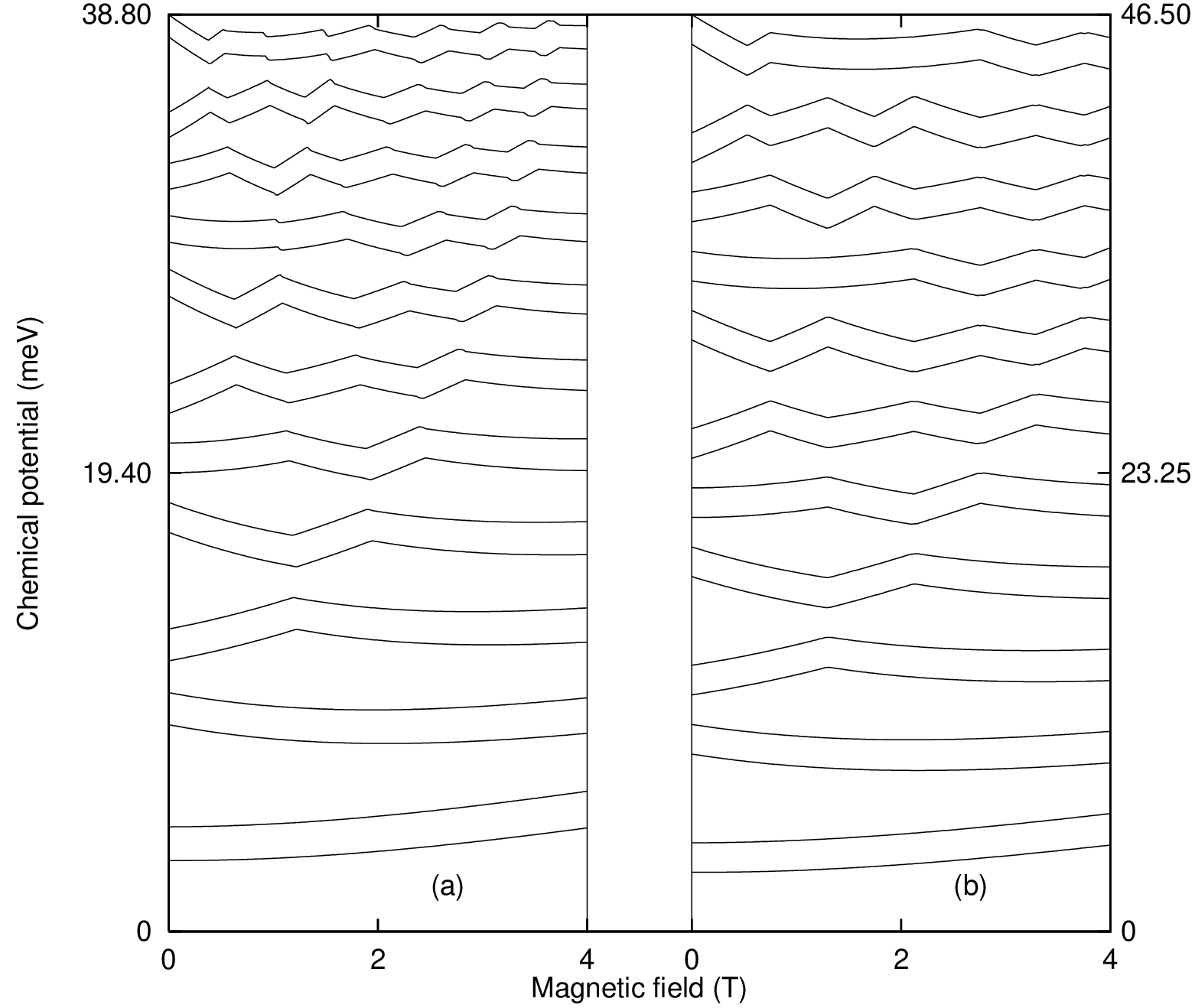}
\begin{small}
{\small 
FIG.~1 Chemical potential $\mu _{N}$ versus magnetic field $B$ for 
$N  \le  22 $: $(a)$ HI model for $\hbar U /(m_*\omega_0) = 0.07$ meV;  
$(b)$ CI model. Parameter values are $\hbar \omega_0 = 3$ meV, 
$\hbar\omega_c/B=1.63$ meV/T, $V_0 =1.5$ meV, $g_* = -0.03 $. }
\end{small}
\end{figure}
\vspace*{0.5cm}
Henceforth we need only consider 
the lowest energy c.m. wave function and for the time being we thus set 
$\Psi_{\rm cm}=\exp\{-\omega\sum_{i,j=1}^N{\bfm r}_i\cdot{\bfm r}_j/(2N)\}$, 
for which one has $E_{\rm cm}=\omega$ and $L_3^{\rm cm}=0$. If one looks for 
eigenstates of Eq.~(\ref{calogero_mod}) in the form 
$\Psi_r = \Phi\Psi_0$, where  
$\Psi_0=\exp\{-\Omega \sum_{i<j}\vert {\bfm r}_i-{\bfm r}_j\vert^2/(2N)\}$, 
and employes holomorphic coordinates $z_i=x_{1,i}+ix_{2,i}$, and 
${\bar z}_i=x_{1,i}-ix_{2,i}$, one gets that the unknown function $\Phi$ must 
satisfy the couple of equations 
\begin{eqnarray}
& - & \sum_{i=1}^N  (2{\bar\partial}_i{\partial}_i - \Omega_+ z_i{\partial}_i -
\Omega_- {\bar z}_i{\bar\partial}_i) \Phi 
 + {{\omega}_c\over2}g_*S_3 \Phi  \nonumber\\
& = & (E-{\cal E}_0-E_{\rm ch})\Phi, 
\qquad
{\bfm P}\Phi=0, 
\eqnum{3$a$,$b$}
\label{hermite-gen}
\end{eqnarray}
where ${\cal E}_0=\Omega(N-1)+\omega$ is the zero-point energy, 
$\Omega_{\pm}=\Omega\pm\omega_c/2$, and we have used the shorthand notation 
$\partial_i={\partial/\partial z_i}$, and 
${\bar\partial}_i={\partial/\partial {\bar z}_i}$. Because the function 
$\Psi_0$ is completely symmetric, $\Psi_r$ and $\Phi$ must have same symmetry.
Eq.~(3$a$) is a sum of $N$ separated Hamiltonians 
and its solutions are built up in terms of single-particle orbital wave 
functions $f_{n_1,n_2}$, with eigenvalues  $\varepsilon_{n_1,n_2}$, 
\begin{eqnarray}
f_{n_1,n_2} & = & e^{\Omega{\bar z} z} {\bar\partial}^{n_1}{\partial}^{n_2}
                 e^{-\Omega{\bar z} z}, 
\nonumber\\
\varepsilon_{n_1,n_2} & = & \Omega (n_1 + n_2) + {{\omega_c}\over2}
(n_1-n_2), 
\eqnum{4} 
\label{levels}
\end{eqnarray} 
where $n_1$ and $n_2$ are non negative integers. Apparently, the original 
interacting problem reduces to a free problem and the main effect of the 
interaction seems just related to a redefinition of the effective frequency 
$\omega\to \Omega$. However, Eq.~(3$b$) is highly non trivial and 
already the two-particle problem shows that it can be satisfied in general 
only by taking linear combinations of degenerate $N$-particle solutions. 
Moreover, the frequency $\Omega=\sqrt{\omega^2-NU}$ depends on $N$, so 
that energy differences like $E(N)-E(N-1)$ cannot be analyzed in terms 
of one-particle levels. In these respects the system is correlated. 

Fortunately, as expected on general ground, one can easily check that the 
ordinary Slater determinant $Z$ obtained by filling the $N$ lowest states 
(\ref{levels}) (so that $S_3=N/2$) is a solution of {\em both} 
Eqs.~(\ref{hermite-gen}). Fig.~(2) shows some typical shape of the set of 
occupied levels. 
By increasing $B$ one observes a depletion of the levels in the $SE$ side and 
an extra filling in the $NW$ side, because the contribution 
$H_L=\omega_c L_z/2$ favours decrease of the orbital angular momentum. 
Noticing that the functions $f_{n_1,n_2}$ are polynomials and using 
standard properties of determinants under column addition and multiplication, 
it is easy to see that for any set of the form depicted in Fig.~(2) 
the Slater determinant can be rearranged into the compact form 
$Z=\det[{z}_{i}^{n_1}{\bar z}_{i}^{n_2}]$. This expression is invariant 
under translation $z_i\to z_i+z_0$, $\forall i$, so that 
${\bfm P}Z=0$ as requested. Introducing the spin index $\sigma=1,2$ for up 
and down electrons, respectively, it follows that the GS wave function 
$\Psi_{\rm GS}$ of $H_B$ in the spin sector $S_3=(M_1-M_2)/2$, $N=M_1+M_2$, 
is obtained, up to a normalization constant, by antisymmetrizing the 
product of the spin-up and spin-down wave functions. Denoting ${\cal A}$ 
 the antisymmetrization, we formally have 
\begin{eqnarray}
\Psi_{\rm GS} = \exp\{-{1\over2}&&\sum_{i,j=1}^N{\bar z}_{i}\Gamma_{ij}z_{j}\}
\nonumber\\
\times && {\cal A}\left\{\prod_{\sigma=1}^2 \det[z^{n_1^{\sigma}}_{i_{\sigma}}
               {\bar z}^{n_2^{\sigma}}_{i_{\sigma}}]\chi_{\sigma}\right\}, 
\eqnum{5} 
\label{ground-state}
\end{eqnarray}
where $z_{i_{\sigma}}$, with $i_{1}=1,\ldots,M_1$ and $i_{2}=M_1+1,\ldots,N$, 
are the coordinates of spin-up and spin-down electrons, respectively, 
$\chi_{1}$, $\chi_{2}$, are the totally symmetric spin functions of 
spin-up and spin-down electrons, 
and $(n_1^{\sigma},n_2^{\sigma})$ are the labels of the (lowest 
energy) levels occupied by spin-$\sigma$ electrons. The matrix 
$\Gamma_{ij} =  \Omega \delta_{ij}+N^{-1}(\omega-\Omega){\cal J}_{ij}$ 
is the square root of $\Lambda_{ij}$ and enters Eq.~(\ref{ground-state}) 
upon combination of $\Psi_0$ with the c.m. GS wave function.

For zero magnetic field the GS is obtained by filling consecutively 
the levels along successive diagonals $n_1+n_2 =k$, $ k=0,1,...$ with two 
electrons of opposite spin, so that $\Psi_{\rm GS}$ has $S_z=0$ or 
$S_z=\pm 1/2$. Completion of a diagonal corresponds to one more shell filled. 
When the magnetic field is turned on, for a wide range of $B$ the GS remains 
in the lowest spin sector due to the smallness of $g_*$, whereas as
previously seen the shape of the Fermi sea modifies in order to lower the
orbital angular momentum.
\begin{figure}[m]
\null\vskip -0.truecm
\epsfxsize=7.7cm\epsfbox{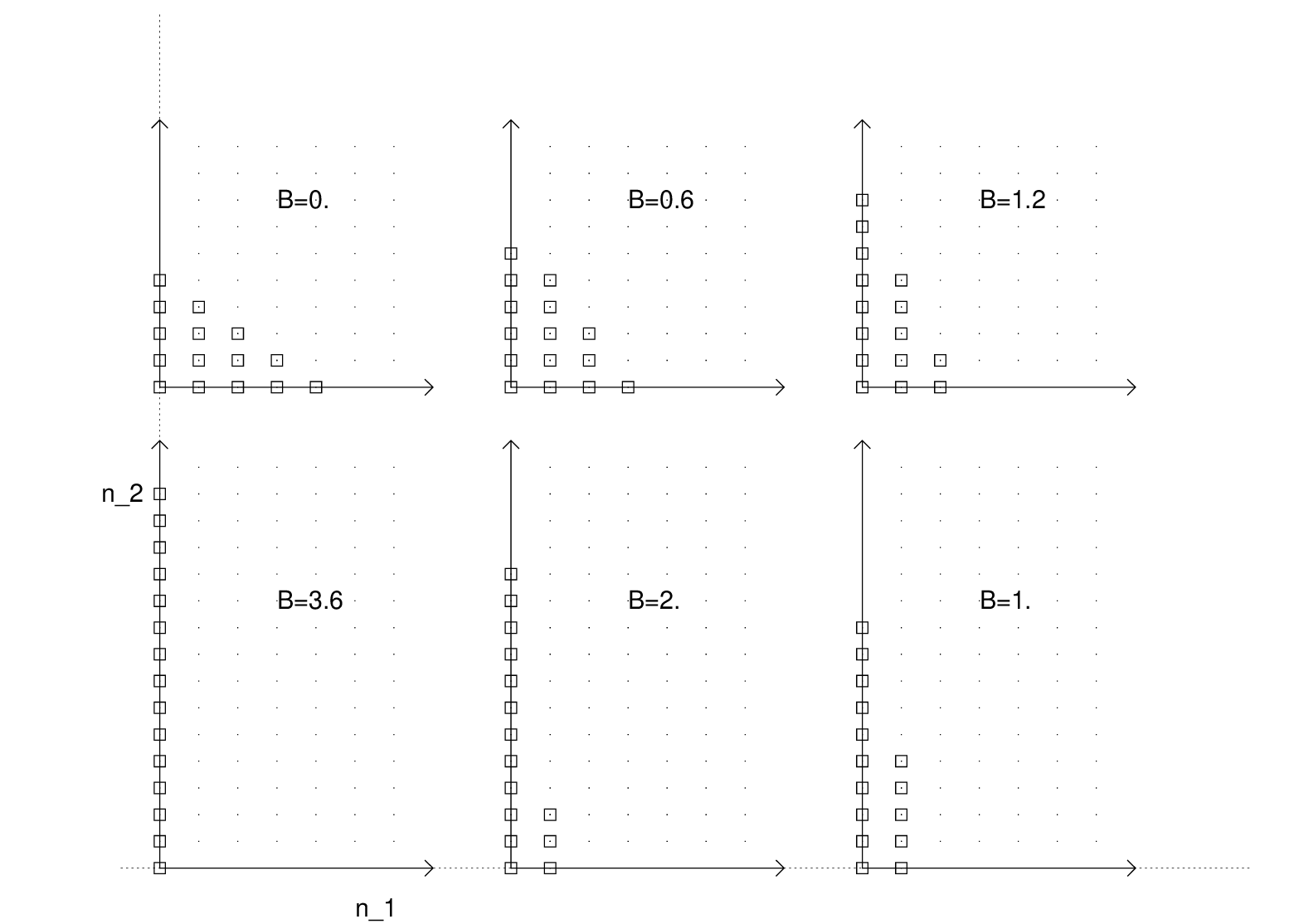}
{\small 
FIG.~2 Occupied single-particle quantum numbers $(n_1,n_2)$ of 
$15$ spinless fermions for six different values of the magnetic field $B$ (T). 
Parameter values are the same as in Fig.~(1a). }
\end{figure}
\vspace*{0.5cm}
The lowest attainable value of the orbital angular momentum 
$L^{\rm min}_3=-\sum_{\sigma}M_{\sigma}(M_{\sigma}-1)/2$ is obtained  
by filling the levels
$(n_1^{\sigma},n_2^{\sigma })=(0,m_{\sigma })$, where
$m_{\sigma}=0,1,\ldots,M_{\sigma}-1$. 
In this case the Vandermonde  determinant  
$Z =\det[{\bar z}_i^m] = \prod_{i<j}({\bar z}_j-{\bar z}_i)$ allows a closed 
expression for the GS wave function $\Psi_{\rm GS}^L$. 
Denoting $\{{\bfm r}_i,\sigma_i\}$ orbital and spin coordinates of the $i$-th 
electron, Eq.~(\ref{ground-state}) gives the Laughlin-like\cite{johnson} state
\begin{eqnarray}
\Psi_{\rm GS}^L = 
\exp\{-{1\over2}&&\sum_{i,j=1}^N{\bar z}_i\Gamma_{ij}z_j\}
\nonumber\\
\times && \prod_{i<j}({\bar z}_j-{\bar z}_i)^{\delta_{\sigma_i,\sigma_j}} 
e^{i{\pi\over2}{\rm sign}(\sigma_i-\sigma_j)}.
\eqnum{6} 
\label{laughlin}
\end{eqnarray} 
Our results for the chemical potential vs. magnetic field are shown in 
Fig.~(1). The curves are splitted by the charging energy $V_0$ but 
the $N$-dependence of $\Omega$ leads to a sizeable 
reduction of the energy scale of the HI model compared with the CI model. 
This behaviour is in agreement with the experimental pattern of 
Ref.\cite{tarucha} and with the discussion in Ref.\cite{schmidt}. Although 
one cannot  overestimate the model nature of the harmonic 
interaction,   our results indicate that the 
rapid decrease of the confinement strength with increasing electron
number can be due to pure correlation effects. The curves (for $N\ge5$) 
perform oscillations that signal transitions to states of lower $L_3$, and 
drop at the field value where the minimum angular momentum state 
(\ref{laughlin}) becomes energetically favourable. It is easy to see that 
this happens when $2\Omega/\omega_c\le M_1/(M_1-2)$ (with $M_1\ge M_2$). 
In the CI model the curves are perfectly paired because of the spin degeneracy 
of the one-particle levels. For example, the transition to the state of lowest 
$L_3$ takes place when $2\omega/\omega_c = M_1/(M_1-2)$. This relation gives 
the same value of the magnetic field at which the transition takes place both 
for $N=2M$ and for $N=2M-1$. Instead, in the HI model the $N$-dependence of 
$\Omega$ breaks this perfect pairing, in particular for large N where $\Omega$ 
becomes small. While usually the angular momentum of the GS is a 
decreasing function of the number of particles, at the field values where
the paired curves come out of phase one instead finds   
$L_z(2N-1) \le L_z(2N)$. For instance, at 
$B=3.6$ T one has $L_z(21)= -89$, whereas  
$L_z(22)=-88$. Hence, the strong reduction of the escillator 
frequency at large $N$ allows for level crossing even within a single 
pair of particles which are added with opposite spin. This in turn leads
to the oscillations out of phase of the peaks forming a pair, a feature 
which is absent in the CI model. We conclude that, in spite of the
simplicity of the model for including electron-electron correlations, many 
features of the experiment of Ref.\cite{tarucha} are nicely reproduced.  

\vspace*{0.7cm}

We would like to thank G. Faini, R. Fazio, A. Mastellone, and U. Merkt
for useful discussions.  A.A. also acknowledges the University of Udine for 
warm hospitality. This work is partly supported by Consiglio Nazionale delle
Ricerche under Contract $\#$ 115.22594.

\end{document}